\documentstyle[preprint,aps,epsf]{revtex}       
\global\let\epsfloaded=Y 
\begin{document}
\pagestyle{empty}                                
\preprint{
\font\fortssbx=cmssbx10 scaled \magstep2
\hbox to \hsize{
\hfill $
\vtop{
 \hbox{ }}$
}
}
\draft
\vfill
\title{SU(3) and Nonet Breaking Effects in $K_L \to \gamma \gamma$ \\
Induced by $s \to d + 2\mbox{gluon}$ due to Anomaly}
\vfill
\author{
X.-G. He$^{1,2}$, Chao-Shang Huang$^1$, and Xue-Qian Li$^{3,4}$}

\address{
$^1$Institute of Theoretical Physics , Academia
Sinica, Beijing\\
$^2$Department of Physics, National Taiwan University, Taipei\\
$^3$ CCAST (World Laboratory), P.O. Box 8730, Beijing\\
$^4$ Department of Physics, Nankai University, Tianjing, 300071}
%
%
\vfill
\maketitle
\begin{abstract}
In this paper we study the effects of $s\to d +  2\mbox{gluon}$
on $K_L \to \gamma\gamma$ in the Standard Model.
We find that this interaction can induce
new sizeable SU(3) and U(3) nonet breaking effects in
$K_L - \eta, \eta'$ transitions and therefore in $K_L\to \gamma\gamma$
due to large matrix elements of
$\langle \eta(\eta')|\alpha_s G^a_{\mu\nu} \tilde G_a^{\mu\nu}|0 \rangle$
from QCD anomaly.
These new effects play an important role in explaining
the observed value. We also
study the effects of this interaction on the contribution to
$\Delta m_{K_L-K_S}$.
\end{abstract}
%
%
%
%
\pagestyle{plain}

It is well known that contributions from intermediate hadronic state
effect
play an important role in many low energy processes.
Some of the notable examples are
$K_L\rightarrow \gamma\gamma$ \cite{1,2} and
$\Delta m_K = m_{K_L} - m_{K_S}$ \cite{3,4,5,6,7}.
For $K_L\rightarrow \gamma\gamma$, the direct contribution due to
quark level $s\to d \gamma\gamma$ alone
accounts for only a small portion of
the amplitude measured experimentally \cite{2,7}.
For $\Delta m_K$, the direct contribution due to $\Delta S = 2$ 
four quark operator
is again only a fraction of the experimental value depending on the
value of the bag factor $B_K$\cite{3,4}.
A simple method to estimate the contributions from intermediate hadronic 
states is
the pole dominance approximation in which one assumes that
a few low lying resonances saturate the contribution.
The commonly identified resonances in the above two cases are
$\pi^0$, $\eta$ and $\eta'$. Combining with U(3) flavor symmetry, the
$K_L \to \gamma \gamma$ amplitude
can be estimated\cite{2,6}.  If U(3) nonet is a good symmetry,
the calculations are straightforward.
However, not only nonet but also SU(3) are known to be
broken, there are large uncertainties in these calculations.
One should also study that if there are some new contributions in 
the Standard Model (SM) which 
have not been examined so far.
In this paper we show that indeed there is a new contribution to 
to $K_L \to\gamma\gamma$ and $\Delta m_K$. This 
new contribution comes from 
$s \to d + 2 \mbox{gluon}$ induced $K-\eta(\eta')$ transition, and the
intermediate $\eta(\eta')$ subsequently decay into $\gamma \gamma$ or
change to another neutral kaon through the usual $\Delta s = 1$ interaction.
We find that, because of the large QCD anomaly hadronic matrix element 
$\langle \eta(\eta')|\alpha_s G^a_{\mu\nu} \tilde G_a^{\mu\nu}|0\rangle$ 
$(\tilde G^{\mu\nu}_a = \epsilon^{\mu\nu\alpha\beta} G^a_{\alpha\beta}$),
the new contributions are sizeable and also  
induce new sizeable SU(3) and U(3) breaking effects.

The decay amplitude $A_{dir}$ of the 
direct contribution to $K_L \to \gamma\gamma$ from quark level 
interaction $s\to d\gamma\gamma$ in the SM has been
studied before\cite{1,2,7}.
Here we improve the calculations by including QCD
corrections which also serve to set up our notations.
In the SM, $ s\to d \gamma \gamma$ can be generated at
one loop level by exchanging a $W$ boson and quarks with two photons emitted
from
particles in the loop and particles in the external legs.
The QCD corrected effective
Hamiltonian for $s\to d \gamma\gamma$ is given by

\begin{eqnarray}
H_{eff}(s\to d \gamma\gamma) = M_{IR}^{\gamma\gamma} + M_R^{\gamma\gamma},
\end{eqnarray}
where $M_{IR}^{\gamma\gamma}$ is the
irreducible contribution with the two photons
emitted from particles in the loop. $M_R^{\gamma\gamma}$ is the reducible
contribution with at least one photon emitted from
an external $s$ or $d$ quark.

The irreducible contribution
$M_{IR}^{\gamma \gamma}$ is given by\cite{2,7,8}

\begin{eqnarray}
M_{IR}^{\gamma\gamma}&=&-i {16\sqrt{2} \alpha_{em} G_F \over 9\pi} N a_2
\epsilon^{*\mu}(k_2) {1\over 2 k_1\cdot k_2} \sum_{i = u,c,t}
V_{id}^*V_{is} F(x,x_i)
\bar d \gamma^\rho L R_{\mu\nu \rho} s \epsilon^{*\mu} (k_1),
\end{eqnarray}
Here $\epsilon^\mu(k)$ is the photon
polarization vector with momentum $k$,
$L(R) = [1 - (+) \gamma_5]/2$, $N = 3$ is the number of colors,
$a_2 = c_1 + c_2/N$,
$x= 2k_1\cdot k_2/m^2_W$, $x_i = m^2_i/m_W^2$, and
$R_{\mu\nu\rho} = k_{1\nu} \epsilon_{\mu \rho \sigma\lambda} k^\sigma_1
k^\lambda_2 - k_{2\mu} \epsilon_{\nu \rho \sigma \lambda} k_1^\sigma
k_2^\lambda + k_1\cdot k_2 \epsilon_{\mu\nu\rho\sigma} (k_2-k_1)^\sigma$.
The function $F(x,x_i)$ is given by
\begin{eqnarray}
F(x,x_i) = {x_i\over x} \int^1_0 {\ln [1-y(1-y) x/x_i]\over y} dy.
\end{eqnarray}

The reducible contribution $M_R^{\gamma \gamma}$ is given by\cite{7,8}
\begin{eqnarray}
M_R^{\gamma\gamma}&=&{\sqrt{2} \alpha_{em}\over 6 \pi}
 \sum_{i = u,c,t} V_{id}^*V_{is} c_{12}^i
\bar d [({1\over p_d\cdot k_1} - {1\over p_s\cdot k_2})
\sigma_{\mu\beta} \sigma_{\nu\alpha} k^\beta_1 k_2^\alpha
+ 2i ({p_{d\mu}\over p_d\cdot k_1} - {p_{s\mu} \over p_s\cdot k_1})
\sigma_{\nu \beta} k^\beta_2\nonumber\\
& +& ( k_1 \to k_2, k_2 \to k_1;
\mu \to \nu, \nu\to \mu) ](m_d L + m_s R) s
\epsilon^{*\mu}(k_1) \epsilon^{*\nu}(k_2).
\end{eqnarray}

In the above $c_i$ are the Wilson coefficients
defined in the following $\Delta S=-1$ effective Hamiltonian\cite{9}

\begin{eqnarray}
&&H_{eff}(\Delta S =-1) = {4G_F\over \sqrt{2}}
[V_{qd}^*V_{qs}(c_1 O_1 +c_2 O_2)
- \sum_{k}\sum_{i = u,c,t} V_{id}^* V_{is} (c_{k}^i O_{k})],
\end{eqnarray}
where the summation over $k$ is on all possible operators, four quark operators,
quark-photon and quark-gluon operators, which are
defined in Ref.\cite{9}. The operators directly relevant to our calculations
to the leading order are

\begin{eqnarray}
&&O_1 = \bar q \gamma_\mu L q \bar d \gamma^\mu L s,\;\;
O_2 = \bar d \gamma_\mu L q \bar q \gamma^\mu L s,\nonumber\\
&&O_{7\gamma} = {e \over 16 \pi^2} \bar d \sigma_{\mu\nu} F^{\mu\nu}
(m_d L + m_s R)s,\nonumber\\
&&
O_{8G} = {g_s\over 16 \pi^2} \bar d \sigma_{\mu\nu} T^a G_a^{\mu\nu}
(m_d L + m_s R)s,
\end{eqnarray}
where $G^{\mu\nu}_a$ and $F^{\mu\nu}$ are the gluon and photon field
strengths. Here we have also written down the operator $O_{8G}$ which
is needed for the study of $s \to d gg$.

To obtain the amplitude $A_{dir}$ 
for $K_L \to \gamma \gamma$ from the effective
Hamiltonian $H_{eff}(s\to d \gamma\gamma)$, one needs to bind the $d$ and $s$
quarks to form a kaon which involves long distance non-perturbative 
QCD effects. This effect cannot be calculated at present and is usually
parameterized by a decay constant $f_K$ as,
$\langle 0| \bar d \gamma^\mu \gamma_5 s | \bar K^0\rangle = - i f_K P_K^\mu$ 
with $f_K$ determined from data.
We have,

\begin{eqnarray}
&&A_{dir}(\bar K^0 \to \gamma \gamma)
= \langle \gamma \gamma |H_{eff}(O_{7\gamma}) | \bar K^0\rangle \nonumber\\ 
&&= {2\sqrt{2} \alpha_{em} G_F\over 9 \pi}f_K [
i(N a_2 V_{ud}^*V_{us} + 3\xi c_{7\gamma}^t V_{td}^*V_{ts})
F_{\mu\nu} \tilde F^{\mu\nu} 
+ 3\xi c_{7\gamma}^t V_{td}^* V_{ts} F_{\mu\nu} F^{\mu\nu}],
\end{eqnarray}
where $H_{eff}(O_{7\gamma})$ indicates the term proportional
to $O_{7\gamma}$ in 
the effective Hamiltonian of eq. (5). 
$\tilde F_{\mu\nu} = (1/2) \epsilon_{\mu\nu\alpha \beta} F^{\alpha\beta}$.

In obtaining the above result, we have used the
fact that $F(x,x_{c,t}) \approx -1/2$ (large $x_{c,t}/x$, and
$F(x,x_u) \approx 0$ (small $x_u/x$).
We also neglected small contributions from
$c^{u,c}_{7\gamma}$ which are proportional to $x_{u,c}$\cite{10}, but have kept
$c_{7\gamma}^t$ which is  $-0.3$ in the SM.

The parameter $\xi$ is an average value of the quantity, $\kappa =
-(m^2_K/ 16) (1/ p_d\cdot k_1 - 1/ p_s\cdot k_2 + 1/ p_d\cdot k_2
- 1/ p_s\cdot k_2)$. If one assumes that the $d$ and $s$ quarks
share equally the kaon momentum, then $\xi = 1$\cite{2}. We have
also estimated $\xi$ by calculating the quantity $<0 | \kappa \bar
d (1+\gamma_5) s |\bar K^0>$ using perturbative QCD method and
appropriate distribution amplitude of quarks in the kaon\cite{11}.
This approach also obtains a value of order one for $\xi$. One
should be aware that the applicability of pQCD may not be a good
one here. However, we find that contribution related to $\xi$ is
not important as long as $\xi$ is of order one. That is, the
precise value of $\xi$ is not important here and we will use $\xi$
to be one in our later discussions.

To estimate the irreducible contribution, one needs to know the
quantity $a_2 = c_1 + c_2/N$. Without QCD corrections, $c_1=0$ and
$c_2=1$. This gives a $a_2=1/3$. With QCD corrections the value
for $a_2$ will be altered. The leading and next-leading order
corrections to $c_i$ have been calculated\cite{9}. The values of
$c_i$ depend on the renormalization scale $\mu$. Since one does
not know precisely where is the matching scale $\mu$, this causes
uncertainty in $a_2$. For example at the leading order, $a_2 =
-0.27$ at $\mu \approx 1$ GeV, while at $\mu = 1.3$ GeV, $a_2 =
-0.17$ with $\Lambda_{\overline{MS}} = 325$ MeV. At the next
leading order the dependences on $\mu$ for each of the $c_1$ and
$c_2$ are reduced, but leaves $a_2$ still sensitive to $\mu$. For
example, in the NDR scheme, for $\Lambda_{\overline{MS}} = 325$
MeV, $a_2$ is -0.08 and -0.1 at $\mu = 1.0$ GeV and $\mu = 1.3$
GeV, respectively. Allowing the QCD parameter
$\Lambda_{\overline{MS}}$ to vary within the allowed range $215
\sim 435$ MeV, $a_2$ can vary in the range $-0.1 \sim -0.35$
depending whether NDR or HV scheme is used\cite{9}. That is, the
value of $a_2$ is not well determined even from the next leading
order perturbative calculations. When all effects, perturbative
and non-perturbative, are correctly treated, the final physical
observables will not depend on the renormalization scale $\mu$.
Unfortunately, such a calculation is not possible at present. The
parameter $a_2$ behaves similarly to the one in hadronic $B$ and
$D$ decays. In both $D$ and $B$ decays, the parameter $a_2$
determined from data ($|a_2|\sim (0.2 \sim 0.5$)) is very
different from factorization value by inserting $c_{1,2}$ at
relevant scale in the expression for $a_2$\cite{12}. One would
expect similar thing happens in kaon decays although the details
may be different. To take into account uncertainties in
theoretical calculations of $a_2$, we will treat it as a free
parameter and allow it to vary in the range of $-0.5 \sim 0.5$.
One can also turn the argument around to obtain information about
$a_2$ from $K_L \to \gamma \gamma$ data.

For $\xi$ of order one, and $a_2$ in the range of $-0.5 \sim 0.5$, we find that
the dominant direct $\bar K^0 \to \gamma \gamma$
amplitude is from the irreducible contribution. We have

\begin{eqnarray}
&&A_{dir}(K_L \to \gamma \gamma)= i\tilde A_{dir}  {1\over 2}
F_{\mu\nu} \tilde F^{\mu\nu},\nonumber\\
&&\tilde A_{dir} =
{8\alpha_{em} G_F \over 9 \pi} f_K N a_2 Re(V_{ud}^*V_{us}).
\end{eqnarray}

Using $V_{ud} = 0.9735$ and $V_{us} =0.2196$ and $f_K = 1.27 f_\pi$\cite{13},
we obtain,

\begin{eqnarray}
\tilde A_{dir} = 2.54\times 10^{-12} a_2 \mbox{MeV}^{-1}.
\end{eqnarray}
For $|a_2| = 0.5$, it is only about 35\% of the experimental value
of $3.5\times 10^{-12}$ MeV$^{-1}$\cite{13}. Without QCD
corrections $a_2 = 1/3$, $\tilde A_{dir}$ is about
24\% of the total amplitude. There must be some other
contributions to this process. These effects may come from 
contributions with intermediate hadronic states or even
contributions from new physics beyond the SM. If one has a good
understanding of all SM contributions, one can have a
detailed study of new physics beyond the SM. It is probably too
early to say that new physics is needed here due to large
uncertainties in possible hadronic intermediate contributions. Therefore
we will work within the SM and see how
contribution from hadronic intermediate states can affect the results.

Several analyses have been carried out using pole model with
$\pi^0$, $\eta$ and $\eta'$ poles to calculate the hadronic 
intermediate contribution.
In this model, the amplitude $A_{had}$ from exchange of
intermediate hadronic states 
is given by\cite{6}
\begin{eqnarray}
\tilde A_{had} &=& \tilde A(\pi^0\rightarrow \gamma\gamma)
{\langle \pi^0|H_W| K_L\rangle \over m_K^2-m_\pi^2} \nonumber\\
& \times&\left[ 1 + {m_K^2-m_\pi^2\over m_K^2 -m_\eta^2}
{\tilde A(\eta\rightarrow \gamma\gamma)\over \tilde
A(\pi^0\rightarrow \gamma\gamma)}
\left( {1+\delta \over \sqrt{3}} \cos\theta +
{2\sqrt{2}\over \sqrt{3}} \rho \sin\theta\right)
 \right. \nonumber\\
& +&\left . {m_K^2-m_\pi^2\over m_K^2-m_{\eta'}^2}
{\tilde A(\eta'\rightarrow \gamma\gamma)\over
\tilde A(\pi^0\rightarrow \gamma\gamma)}
\left({1+\delta\over \sqrt{3}} \sin\theta - 
{2\sqrt{2}\over \sqrt{3}} \rho \cos\theta\right)
 \right],
\end{eqnarray}
where $\theta$ is the $\eta-\eta'$ mixing angle,
$\delta$ is the SU(3) breaking parameter \cite{6}.
The parameter $\rho$ parameterizes U(3) nonet breaking effect
and is defined as
\begin{eqnarray}
\rho = -\sqrt{{3\over 8}}{<\eta_1|H_W|K^0>\over <\pi^0|H_W|K^0>}.
\end{eqnarray}
In the nonet limit $\rho = 1$.
Chiral Lagrangian analysis gives
$\langle \pi^0|H_W| K_L\rangle = 1.4 \times 10^{-7} m_K^2$ \cite{6}.
Using experimental values for $\pi^0, \eta,\eta' \to \gamma\gamma$, 
$\tilde A_{had}$ can be estimated.

The above contributions can be viewed as obtained by $A_{had} 
= \sum_i \langle \gamma \gamma |i\rangle \langle i 
|\tilde H_{eff} (\Delta S = -1)|K_L\rangle$ with $i = \pi^0,\; 
\eta,\;\eta'$ in the pole model approximation. Here 
$\tilde H_{eff}(\Delta S = -1)$ is the full $\Delta S = -1$ 
effective Lagrangian with $O_{7\gamma}$ term removed since it has been
counted as the contribution to $A_{dir}$. Therefore $A_{dir}$ and
$A_{had}$ are contributions from different sources. In previous calculations the
contributions for $A_{had}$ from $O_{8G}$ were not considered\cite{2,6}. We now
study in detail the effect of this interaction on $K_L \to \gamma \gamma$. 

At the quark-gluon level, $O_{8G}$ induces $s\to d + gg$.
To obtain $A_{had}$, 
one needs to estimate the contribution from $s\to d + gg$ to $\bar K^0 - \eta,
\eta'$ through $gg \to \eta, \eta'$.
The effective Hamiltonian $M^{gg}_{IR,R}$ for $s\to d gg$, with
color singlet $\bar d s$ bi-spinor product,
can be obtained by some simple replacements from $M^{\gamma\gamma}_{IR,R}$.
To obtain $M_{IR, R}^{gg}$ one first replaces
the photon polarization vectors $\epsilon^{\mu}(k_1) \epsilon^\nu(k_2)$
by the gluon polarization vector $\epsilon^\mu_a(k_1) \epsilon^\nu_a(k_2)$
with the color index $a$ summed over. Then
one replaces $\alpha_{em}$ by
$\alpha_s (9/4)/(2N)$ and $\alpha_{em} c_{12}^i$ by $\alpha_s c^i_{8G}/(2N)$
for $M_{IR}^{gg}$ and $M_R^{gg}$, respectively\cite{7}. The factor
$1/(2N)$ comes from picking up the color singlet part.

Similar to the procedure in obtaining the amplitude $A_{dir}$ for
$K_L \to \gamma \gamma$,  one can obtain the amplitude for $K_L \to gg$.
We find that with $\xi$ of order one, $a_2$ in the range of $-0.5\sim 0.5$
and $c_{8G}^t \approx -0.15$ as given in the SM, the irreducible
contribution, again, dominates the amplitude. We have

\begin{eqnarray}
A(K_L \to gg) =
{1\over 2 N} {2\alpha_s G_F \over \pi}
f_K N a_2 Re(V_{ud}^*V_{us}) i {1\over 2}
G^a_{\mu\nu} \tilde G_a^{\mu\nu}.
\end{eqnarray}

The above interaction can induce large $K_L - \eta, \eta'$ transitions
and therefore contribution to $K_L \to \gamma\gamma$,
because QCD can induce large matrix elements for
$\langle \eta(\eta')|\alpha_s G^a_{\mu\nu}\tilde G^{\mu\nu}_a|0\rangle$.

QCD anomaly implies that the divergence of the singlet
current, $a^1_\mu = \bar u \gamma_\mu \gamma_5 u + \bar d \gamma_\mu \gamma_5 d
+ \bar s \gamma_\mu \gamma_5 s$, is not zero in the limit of zero
quark masses, and is given by

\begin{eqnarray}
\langle \eta(\eta')|\partial^\mu a^1_\mu|0\rangle &=&
\langle \eta(\eta')|2i(m_u \bar u \gamma_5 u + m_d \bar d \gamma_5 d
+m_s \bar s \gamma_5 s)|0\rangle\nonumber\\
& -&
\langle \eta(\eta')|{3\alpha_s\over 4 \pi} G^a_{\mu\nu} \tilde G^{\mu\nu}_a|0\rangle.
\end{eqnarray}
While for the octet current,
$a^8_\mu = \bar u \gamma_\mu \gamma_5 u + \bar d \gamma_\mu \gamma_5 d
- 2\bar s \gamma_\mu \gamma_5 s$, one obtains\cite{14}

\begin{eqnarray}
&&\langle \eta(\eta')|\partial^\mu a^8_\mu|0\rangle  = \langle \eta(\eta')|
2i(m_u \bar u \gamma_5 u + m_d \bar d \gamma_5 d
-2m_s \bar s \gamma_5 s)|0\rangle .
\end{eqnarray}
Since $m_{u,d}$ are much smaller than $m_s$, one can neglect
terms proportional to $m_{u,d}$. One then obtains

\begin{eqnarray}
&&\langle \eta'(p)|{3\alpha_s \over 4\pi} G_{\mu\nu}^a \tilde G^{\mu\nu}_a
|0\rangle
= \sqrt{3\over 2} ( \sqrt{2} f_1 \cos\theta + f_8 \sin \theta)p^2,\nonumber\\
&&\langle \eta(p)|{3\alpha_s \over 4\pi} G_{\mu\nu}^a \tilde G^{\mu\nu}_a
|0\rangle
= \sqrt{3\over 2} ( -\sqrt{2} f_1 \sin\theta + f_8 \cos \theta)p^2,
\end{eqnarray}
where $f_{1,8}$ are the singlet and octet pseudo-scalar decay constants.

If there is no $\eta-\eta'$ mixing and all quark masses are equal,
the $gg$ state being a flavor singlet can only have transition to
$\eta_1$. Because the $\eta -\eta'$
mixing and the different quark masses, both
U(3) nonet and SU(3) symmetries are broken. The $K_L \to \eta, \eta'$
transitions induced by $s\to d gg$ will induce nonet and SU(3)
breaking in the total amplitude $\tilde A^{total }$. Normalizing the
signs of each contributions to theoretical calculations, we finally obtain

\begin{eqnarray}
\tilde A^{total} &=& \tilde A_{dir} +
\tilde A(\pi^0\rightarrow \gamma\gamma)
{\langle \pi^0|H_W| K_L\rangle \over m_K^2-m_\pi^2} \nonumber\\
& \times& \left[ 1 + {m_K^2-m_\pi^2\over m_K^2 -m_\eta^2}
{\tilde A(\eta\rightarrow \gamma\gamma)\over 
\tilde A(\pi^0\rightarrow \gamma\gamma)}
\left( {1+\delta +\delta^{gg} \over \sqrt{3}} \cos\theta +
{2\sqrt{2}\over \sqrt{3}} (\rho+r^{gg}) \sin\theta\right)
 \right. \nonumber\\
&+ &\left . {m_K^2-m_\pi^2\over m_K^2-m_{\eta'}^2}
{\tilde A(\eta'\rightarrow \gamma\gamma)\over \tilde
A(\pi^0\rightarrow \gamma\gamma)}
\left({1+\delta+\delta^{gg} \over \sqrt{3}} \sin\theta -
{2\sqrt{2}\over \sqrt{3}} (\rho+r^{gg}) \cos\theta\right)
 \right],
\end{eqnarray}
where $\delta^{gg}$ and $r^{gg}$ are the SU(3) and nonet breaking
induced by the $s\to d gg$ interaction. They are given by

\begin{eqnarray}
&&\delta^{gg} = -\sqrt{2}
f_Kf_8 m^2_K {G_F Re(V_{ud}^*V_{us}) \over \langle \pi^0| H_W| K_L\rangle
}a_2,\nonumber\\
&&r^{gg} = - {f_1\over 2 f_8} \delta^{gg}.
\end{eqnarray}

We find

\begin{eqnarray}
\delta^{gg} = 0.96{f_8\over f_K} a_2,
\;\;\;\;r^{gg} = -0.48 {f_1\over f_K} a_2.
\end{eqnarray}
We see that the corrections can be sizeable and can not be neglected.

We now provide some details for numerical calculations. 
There are several parameters
involved in $\tilde A_{had}$, the mixing angle $\theta$,
the decay constants $f_{1,8}$, the SU(3) and U(3) nonet breaking parameters
$\delta$ and $\rho$, and the parameter $a_2$. Chiral perturbation
calculations and fitting data not involving $K_L \to \gamma\gamma$
have obtained $\theta \approx -20^\circ$, $\delta \approx 0.17$,
$f_8 \approx 1.28 f_\pi$ and $f_1\approx 1.10f_\pi$\cite{15}.
We will use these values for these
parameters in the calculation of $K_L \to \gamma\gamma$.
There is not a reliable estimate for the parameter
$\rho$. Since we are interested to see how the new $s\to d gg$ interaction
induces U(3) nonet breaking effect, we will take $\rho=1$ and attribute
nonet breaking solely to $r^{gg}$. As have been
discussed $s\to d gg$ also induce
SU(3) breaking effect. This effect was not included in other fittings. We
therefore should include this new SU(3) breaking effect also.

Without the $s\to d gg$ effect, we find that the amplitude $\tilde A^{total}$
is equal to $5.5(1 + 0.46a_2)\times 10^{-12} $ MeV$^{-1}$ which is
considerably larger than
the experimental value $3.5\times 10^{-12}$ MeV$^{-1}$\cite{13}
for $|a_2| < 0.5$.
With the new effect,
we find

\begin{eqnarray}
\tilde A^{total} = 5.5 (1+2.14 a_2)\times 10^{-12} \mbox{MeV}^{-1}.
\end{eqnarray}
To reproduce the central experimental value, $a_2$ is required to be $-0.17$
which is a reasonable value to have.

The detailed numerical results depend on
several parameters. Even with other parameters fixed, one can introduce
also a phase to $a_2$. To fit the $K_L\to \gamma \gamma$ data,
the values for the magnitude and phase of $a_2$ can vary. We, however, would
like to emphasize that the new effect discussed can play an important role
in $K_L \to \gamma \gamma$ independent of the details.

The new contributions for $K_L - \eta (\eta')$ transitions also induce
new hadronic intermediate state effect to the $K_L$ and $K_S$ mass difference
parameter $\, {\rm Re} (M_{12})$ in the pole dominance approximation.
We find\cite{6}
\begin{eqnarray}
2m_K \, {\rm Re} (M_{12}) &=& {|\langle \pi^0|H_W|K^0\rangle|^2\over m_K^2-m_\pi^2}
\nonumber\\
&\times&
\left[ 1+ {m_K^2-m_\pi^2\over m_K^2-m_\eta^2} \left( {1+\delta+\delta^{gg}
\over \sqrt{3}} \mbox{cos}\theta
+{2\sqrt{2}\over \sqrt{3}}(\rho+r^{gg}) \mbox{sin}\theta\right)^2 \right .\nonumber\\
&+& \left .
{m_K^2-m_\pi^2\over m_K^2-m_{\eta'}^2} \left( {1+\delta+\delta^{gg}\over \sqrt{3}} \mbox{sin}\theta - {2\sqrt{2}
\over \sqrt{3}} (\rho+r^{gg}) \mbox{cos}\theta \right)^2 \right].
\end{eqnarray}

Without the new effects, the above would lead
to $\Delta m_K =-0.5\times 10^{-12} $ MeV
which is a non-negligible portion of the
experimental value of $3.5\times 10^{-12}$ MeV. With the new effects and
$a_2 = - 0.17$ as determined from $K_L \to \gamma\gamma$, the contribution to
$\Delta m_K$ is $-0.9 \times 10^{-12}$ MeV,
and again it can not be neglected. The new effect
in $K_L \to \pi^0,\;\eta,\;\eta'$ transitions can have sizeable
contribution to $\Delta m_K$.

The $s \rightarrow d gg$ process can also induce $K_L$-glueball mixing,
which would also affect $K_L\rightarrow \gamma\gamma$
and $\Delta m_{S-L}$, as pointed out in Ref. \cite{7} where
a light glueball mass $1.4$ GeV was used.
Recent lattice calculations indicate that the pseudo-scalar
glueball mass is about $2.3$ GeV \cite{16}.
With such a large mass the glueball-$\eta (\eta')$ mixing contribution
should be small and therefore the effects are
smaller than effects discussed earlier.

In conclusion we have evaluated additional contributions to
$K_L\rightarrow \eta (\eta')$ transitions from $s\rightarrow  d gg$
in the Standard Model.
These transitions induce new sizeable SU(3) and U(3)
breaking effects and have significant
effects on contributions to $K_L\rightarrow \gamma\gamma$
and $\Delta m_K$.

\noindent
{\bf\large Acknowledgments}

The work of XGH
was supported in part by
National Science Council under grants NSC
91-2112-M-002-42,
and in part by the Ministry of
Education Academic Excellence Project 89-N-FA01-1-4-3.
The work of CSH and XQL is
supported in part by National Natural
Science Foundation. XGH would like to thank the hospitality of
Institute for Theoretical Physics in Beijing where part of this
work was carried out. He would also like to thank Hai-Yang Cheng
for useful discussions.


\begin{thebibliography}{99}
\bibitem{1} M.K. Gaillard and B.W. Lee, Phys. Rev. {\bf D10}, 897(1974).

\bibitem{2} E. Ma and A. Pramudita, Phys. Rev. {\bf D24}, 2476(1981).

\bibitem{3}L. Wolfenstein, Nucl. Phys. {\bf B160}, 50(1979)

\bibitem{4} G. D'Ambrosio and D. Esprin, Phys. Lett. {\bf B175}, 237(1986);
J.L. Goity, Z. Phys. {\bf C34}, 341(1987).

\bibitem{5} L.L. Chau and H.Y Cheng, Phys. Lett. {\bf B195}, 275(1987).

\bibitem{6} J. Donoghue, B. Holstein and Y.C. Lin, Nucl. Phys. {\bf B277}, 651(1986).

\bibitem{7}X.G. He, S. Pakvasa, E.A. Paschos, and Y. L. Wu, Phys. Rev. Lett. {\bf 64}, 1003(1990).

\bibitem{8} C.H. Chang, G.L. Lin and Y.K. Yao, Phys. Lett. {\bf 415}, 395(1997).

\bibitem{9} G. Buchalla, A. Buras and M. Lautenbacher, Rev. Mod. Phys.
{\bf 68}, 1125(1996).

\bibitem{10} X.-G. He and G. Valencia, Phys. Rev. {\bf D61}, 075003(2000).

\bibitem{11} V. Braun and I. Filyanov, Z. Phys. {\bf C44}, 157(1989);
Z. Phys. {\bf C48}, 239(1990).

\bibitem{12}, H.-Y. Cheng e-print hep-ph/0202254;
C.-K. Chua, W.-S. Hou and K.-C. Yang, Phys. Rev. {\bf D65}, 096007(2002);
M. Neubert and A. Petrov Phys. Lett. {\bf B519}, 50(2001);
Zhi-Zhong Xing e-print hep-ph/0107257.

\bibitem{13} Particle Data Group, Eur. Phys. J. {\bf C15}, 1(2000).

\bibitem{14} R. Akhoury and J.-M. Frere, Phys. Lett. {\bf B220}, 258(1989); P. Ball,
J.-M. Frere and M. Tytgat, ibid, {\bf B365}, 367(1996);
X.-G. He, W.-S. Hou and C.-S. Huang, Phys. Lett. {\bf B429}, 99(1998).

\bibitem{15} E. Venugopal and B. Holstein, Phys. Rev. {\bf D57}, 4397(1998);
T. Feldmann and P. Kroll, Eur. Phys. J. {\bf C5}, 327(1998).

\bibitem{16} G.S. Bali et al., UKQCD collaboration, Phys. Lett. {\bf 309}, 378(1993).

\end{thebibliography}
\end{document}